\documentclass[aps,prl,twocolumn,showpacs]{revtex4-1}
\usepackage{amsfonts}
\usepackage{amsmath}
\usepackage{amsthm}
\usepackage{graphics}
\usepackage{graphicx}
\usepackage{subfigure}
\usepackage{amssymb}
\usepackage{graphics}
\usepackage{graphicx}
\usepackage{epstopdf}
\usepackage{color}
\usepackage{subfigure}
\usepackage{pgfplots,tikz}
\usepackage{url}
\usepackage{float}
\usepackage{soul}
\def\x{{\bf x}}

\newcommand\dertt[1]{ \frac{\partial{ #1}}{\partial t} }

\def\Vp{V_{\rm p}}
\def\Mp{M_{\rm p}}
\def\Rp{a_{\rm p}}
\def\Rpbar{\bar{a}_{\rm p}}

\begin{document}

\title{Stokes drift and impurity transport in a quantum fluid}
\author{Umberto Giuriato}
\affiliation{
Universit\'e C\^ote d'Azur, Observatoire de la C\^ote d'Azur, CNRS, Laboratoire Lagrange, Bd de l'Observatoire, CS 34229, 06304 Nice cedex 4, France.}
\author{Giorgio Krstulovic}
\affiliation{
Universit\'e C\^ote d'Azur, Observatoire de la C\^ote d'Azur, CNRS, Laboratoire Lagrange, Bd de l'Observatoire, CS 34229, 06304 Nice cedex 4, France.}
\author{Miguel Onorato}
\affiliation{Dipartimento di Fisica, Universit\`a degli Studi di Torino and INFN, Sezione di Torino, Via Pietro Giuria 1, 10126 Torino, Italy}
\author{Davide Proment}
\affiliation{School of Mathematics, University of East Anglia, Norwich Research Park, Norwich, NR4 7TJ, United Kingdom}


\begin{abstract}
Stokes drift is a classical fluid effect in which travelling waves transfer momentum to tracers of the fluid, resulting in a non-zero drift velocity in the direction of the incoming wave.
This effect is the driving mechanism allowing particles, i.e. impurities, to be transported by the flow; in a classical (viscous) fluid this happens usually due to the presence of viscous drag forces. 
Because of the eventual absence of viscosity in quantum fluids, impurities are driven by inertial effects and pressure gradients only.
We present theoretical predictions of a Stokes drift analogous in quantum fluids for classical impurities obtained using multi-time analytical asymptotic expansions.
We find that, at the leading order, the drift direction and amplitude depend on the initial impurity position with respect to the wave phase; at the second order, dominant after averaging over initial conditions, our theoretical model recovers the classical Stokes drift but with a coefficient that depends on the relative particle-fluid density ratio.
Numerical simulations of a two-dimensional Gross-Piteaveskii equation coupled with a classical impurity corroborate our findings.
Our predictions are experimentally testable, for instance, using fluids of light obtained in photorefractive crystals.
\end{abstract}
\maketitle

The transport properties caused by waves and, more in general, the interaction of waves with particles has  been a long standing problem in many field of physics. 
A seminal example is given by particles floating under or on the surface of  propagating (incompressible) water waves that experience a velocity drift which is known as the Stokes drift. 
This phenomenon, first described by G.~G. Stokes in 1847 \cite{stokes1880theory}, is related to the intrinsic nonlinearity that characterises the Lagrangian description of a  water particle immersed  in a linear (or nonlinear) Eulerian wave field. 
Water particles move on trajectories that are not closed and, on average, advance in the direction of the propagation of the waves  with a velocity that is one order of magnitude smaller that the  phase velocity.
This drift is important for the mass transfer of any object in a wave field specially in the area of sediment transport \cite{longuet1953mass,besio2004modeling,santamaria2013stokes} and it is responsible of important fluid-mixing processes. 

Acoustic waves do also transport particles:  just like in an  optical wave field, particles in an acoustic  field are affected by the so-called acoustic radiation force that is the result of a transfer of linear momentum from waves to a particle \cite{bruus2012acoustofluidics, toftul2019acoustic,xie2014dynamics,cleckler2012motion}. 
The first calculation of this effect was reported in 1934 by L. King \cite{king1934acoustic} showing that the radiation pressure is always in the direction of propagation of the wave. 
Since then, there has been a  considerable interest in modelling  the particle dynamics that results from averaging over many wave periods, especially because small particles may be used as tracers to visualise the flow.

The transport of particles in quantum fluids exhibiting both normal (viscous and thermal) and superfluid components has been investigated only in the last couple of decades \cite{PhysRevB.71.064514, PhysRevFluids.5.032601}.
Particles move due to the presence of pressure gradients and the drag force caused by the viscous (\emph{normal}) component.
In superfluid liquid helium, density waves have very low amplitude due to the low compressibility of the system and pressure gradients origin mainly from the velocity field produced by quantised vortices.
Particles become trapped into the core of quantised vortices \cite{Sergeev:2009un,giuriatoInteractionActiveParticles2019} and have been used as tracers to probe, for example, the existence of quantised vortex filaments \cite{PhysRevLett.33.280} and vortex reconnections \cite{P.:2008tp}.
Also, note that in the typical liquid helium experiments with particles the normal component cannot be neglected, hence particles also experience a classical Stokes drag.

In the limit of zero temperature, a quantum fluid has no normal component in the flow and particles are driven only by pressure gradients of the superfluid component as viscous dissipation is absent \cite{giuriatoInteractionActiveParticles2019}.
Differently to liquid helium, weakly interacting quantum fluids are highly compressible and both classical \cite{refId0} and quantum \cite{Clark:1965wp} particles (usually called impurities in this context) can move thanks to the interaction with vortices \cite{PhysRevB.63.024510, PhysRevB.97.094507, giuriatoHowTrappedParticles2020} but also due to density waves \cite{giuriatoStochasticMotionFinitesize2021}. 
Examples of such fluids are dilute gaseous Bose-Einstein condensates and quantum fluids of light, both of which can be quantitatively described using the Gross-Pitaevskii (GP) semiclassical model.
In this Letter we present the Stokes drift and impurity transport in such setting.

For simplicity in the analytical predictions and numerical calculations, we consider a classical impurity (that is a classical-like particle having a well-defined position and momentum) whose characteristic size is of the order of the healing length of the system, and analyse how the acceleration caused by a superfluid density wave transports the impurity. 
The quantum fluid is described by a complex field $\psi(\mathbf{x},t)$ and the impurity classical degrees of freedom, position and momentum, are $ \mathbf{q} $ and $ \mathbf{p}=\Mp\dot{\mathbf{q}} $, respectively, given $ \Mp $ the impurity's mass. 
The dynamics of the system is governed by the (GP) equation coupled with a classical Newton's equation for the impurity; they read
\begin{eqnarray}
i\hbar\dertt{\psi} &=& - \frac{\hbar^2}{2m}\nabla^2 \psi + \left( g|\psi|^2-\mu\right)\psi+\Vp(| \x -{\bf q} |)\psi \label{Eq:GPEParticles}, \\ 
\Mp\ddot{\bf q} &=& - \int  \Vp(| \x -{\bf q}|) \nabla|\psi|^2\, \mathrm{d} \x,
\label{Eq:Particle}
\end{eqnarray}
where $m$ is the mass of the fundamental boson of the quantum fluid, $\mu$ is its chemical potential and $g$ is the coupling constant of boson-boson local interaction. 
The potential $ \Vp(0)\gg\mu $ is localized around $\mathbf{q}$ and effectively determines the size of the impurity as its presence induces a complete depletion of the quantum fluid about the position $\mathbf{q}$ up to the characteristic distance $\Rp$ where $ \Vp(\Rp) = \mu $.
The total energy of the system, the quantum fluid mass $M=m\int|\psi|^2\,\mathrm{d}\mathbf{x}$ and the total momentum $\mathbf{P}=\frac{i\hbar}{2}\int(\psi\nabla\psi^*-\psi^*\nabla\psi)\,\mathrm{d}\mathbf{x}+\mathbf{p}$ are conserved quantities. This model has been successfully used to describe the interaction between impurities mediated by the superfluid \cite{shuklaStickingTransitionMinimal2016} and their interaction with quantum vortices and Kelvin waves \cite{giuriatoHowTrappedParticles2020,giuriatoInteractionActiveParticles2019,giuriatoQuantumVortexReconnections2020}.

The system has a hydrodynamical interpretation via the Madelung transformation $\psi(\x)=\sqrt{{\rho(\x)}/{m}}\,e^{i\frac{m}{\hbar}\phi(\x)}$ that maps Eq.\eqref{Eq:GPEParticles} into the continuity and Bernoulli equations of a fluid of density $\rho$ and velocity $\mathbf{v}_\mathrm{s}=\nabla\phi$. 
In absence of the impurity, the GP equation has a simple steady solution corresponding to the uniform state (condensate) $ |\psi_0|=\sqrt{{\rho_0}/{m}}=\sqrt{\mu/g}$.  
If Eq.\eqref{Eq:GPEParticles} is linearised about $\psi_0$, large wavelength waves propagate with the phonon (sound) velocity $c=\sqrt{g\rho_0/m^2}$ and dispersive effects take place at length scales smaller than the healing length $\xi=\sqrt{\hbar^2/2g\rho_0}$. 
Using fluid dynamical variables, wave perturbations of the uniform solution along the $ x $-direction, for instance, are simply 
\begin{equation}
\rho = \rho_0 + A_\rho\cos{(kx-\omega t)},\quad v_\mathrm{w} = \frac{\omega A_\rho}{k\rho_0}\cos{(kx-\omega t)}
  \label{Eq:wave}
  \end{equation}
where we assume $ A_\rho/\rho_0 \ll 1 $ and the angular frequency $ \omega $ is the celebrated Bogoliubov dispersion relation 
\begin{equation}
\omega = c|k|\sqrt{1+\frac{\xi^2k^2}{2}}.
\label{Eq:bogo}
\end{equation}
Please refer to Section I of the Supplemental Material (SM) for detailed derivation.
The wave period and wave length are thus defined as $T = 2\pi/\omega$ and $\lambda=2\pi/k$, respectively.

We integrate numerically Eqs.(\ref{Eq:GPEParticles}-\ref{Eq:Particle}) using the standard pseudo-spectral code FROST \cite{krstulovicHDR}, and setting as initial condition the superposition of a linear wave solution, Eq.\eqref{Eq:wave}, with the ground state solution (obtained numerically by imaginary time evolution) of the impurity immersed in the quantum fluid.
For simplicity we consider only two spatial dimensions with a double periodic rectangular domain of size $1024\xi\times128\xi$, using $1024\times128$ collocation points. 
The potential used to model the impurity is a smoothed hat-function $\Vp(r)=V_0/2\{1-\tanh\left[(r^2 -\eta_a^2)/(4\Delta_a^2)\right]\}$. 
Note that because of the nonlinearity/dispersion balance of the GP system, the quantum fluid density at the impurity boundary takes a distance $\sim \xi$ to heal to the bulk value. 
Thus, we can define an effective particle radius $\Rpbar>\Rp$, estimated by measuring the volume of the displaced fluid 
$\pi \Rpbar^2 = \int (|\psi_0|^2 - |\psi_\mathrm{p}|^2)\,\mathrm{d}\mathbf{x}$, where $\psi_\mathrm{p}$ is the steady state with one impurity. 
We express the non-dimensional impurity mass as $\mathcal{M}=M_\mathrm{p}/M_0$, where $M_0=\rho_0\pi \Rpbar^2$ is the mass of the displaced fluid. 
In all the simulations, we fix the impurity potential $V_0=20\mu$. 
We set the hard-core size to $\Rp=1.5\xi$ and the effective size to $\Rpbar = 3.1\xi$ by choosing $\eta_a = \xi$ and $\Delta_a = 0.75\xi$.
For brevity, we will indicate the impurity position along the wave direction ($ x $-axis) simply as $q\equiv q_x$.

\begin{figure}
  \centering
  \includegraphics[width=1\linewidth]{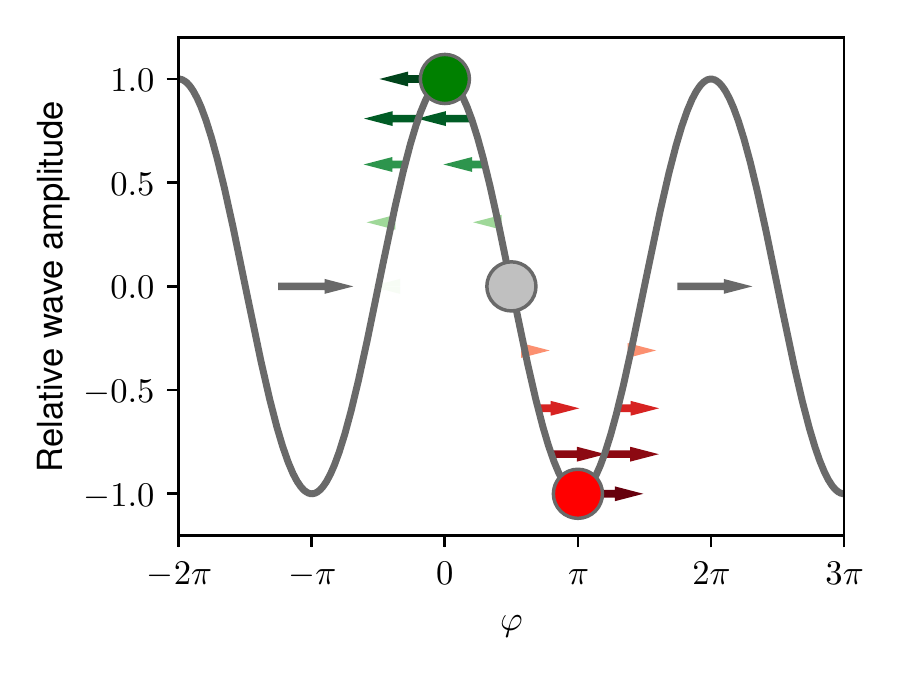}
  \caption{Sketch of the drift. At the leading order, the drift direction and magnitude depends on the initial impurity-wave phase $ \varphi $.}
  \label{Fig:sketch}
\end{figure}

\begin{figure}
  \centering
  \includegraphics[width=1\linewidth]{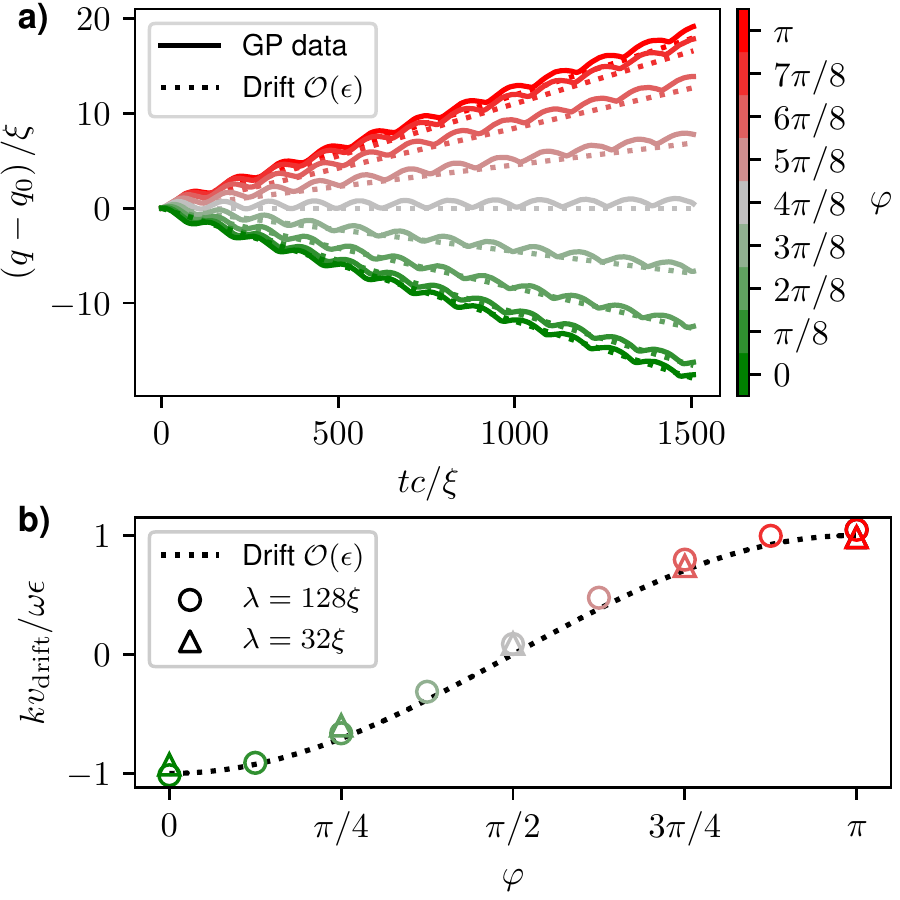}
  \caption{\textbf{a)} Time evolution of the impurity rescaled position (solid lines) for different impurity-wave phases. Dotted lines represent the drift prediction obtained in Eq.\eqref{Eq:drift} at the leading order. \textbf{b)} Rescaled drift versus the impurity-wave phases for waves of wavelength $\lambda=128\xi$ (circles) and $32\xi$ (triangles); the dotted line is the prediction \eqref{Eq:drift} at the leading order.}
  \label{Fig:phase}
\end{figure}

Initially, we place the impurity with zero velocity at different positions $q_0=q(t=0)$ with respect to the phase of the incident wave.
\begin{figure}
\centering
\includegraphics[width=1\linewidth]{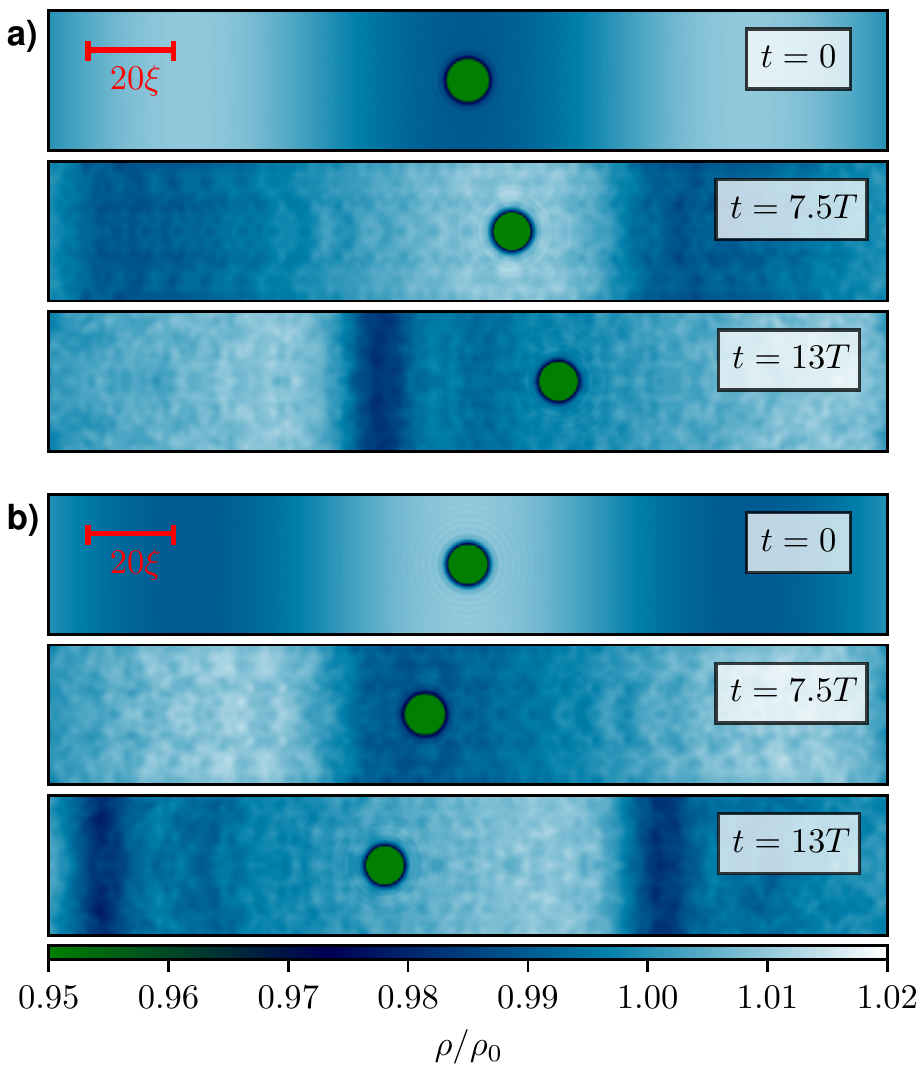}
\caption{
Rescaled quantum fluid density $ \rho/\rho_0 $ at different times in the presence of a density wave of wavelength $\lambda=128\xi$ moving from left to right and an impurity (green dot) of relative mass $\mathcal{M}=0.1$. 
Note that the plots show only a portion of the entire (periodic) computational box. 
\textbf{a)} Wave amplitude $A_\rho/\rho_0=0.01$ and initial position of the impurity in the trough of the wave. 
\textbf{b)} Wave amplitude $A_\rho/\rho_0=0.01$ and initial position of the impurity on the crest of the wave. 
}
\label{Fig:density_evo}
\end{figure}
If the impurity is placed at the wave trough, we observe a drift in the same direction of propagation of the wave, see Fig.\ref{Fig:density_evo}.a); if it is placed a the crest, the impurity moves in the opposite direction, see Fig.\ref{Fig:density_evo}.b). 
Note that the figures show only a fraction of the numerical box, zoomed closed to the impurity.
Our numerical observations are summarized in the sketch of Fig.\ref{Fig:sketch} where we define the initial impurity-wave phase as $\varphi = q_0k$, with the convention that $\varphi=0$ when $q_0$ is at the wave crest.
In order to quantitatively characterize the drift effect, we monitor the impurity displacement as a function of time, for different initial impurity-wave phase $\varphi$s. 
The impurity drift is displayed in Fig.\ref{Fig:phase}.a). 
In Fig.~\ref{Fig:phase}.b), we show the measured drift velocity $v_\mathrm{drift}$, computed averaging the impurity velocity over $13$ wave periods, as a function of the impurity-wave phase and for two different wavelengths of the carrier wave.
As the phase is changed from 0 to $ \pi $, we observe a smooth transition from backward to forward drift.

To explain the change of direction of the impurity drift with respect to the impurity-wave phase, we build an effective minimal model to describe the problem. 
We start by considering that the force acting on the impurity, the right hand side of Eq.\eqref{Eq:Particle}, is nothing but the fluid density gradient convoluted with the impurity potential. 
As formally derived in Section II of the SM, if the impurity size is much smaller than the wavelength $ \lambda $ and neglecting any active effect of the impurity onto the fluid, the impurity dynamics is driven by the effective equation
\begin{equation}
  \ddot{q} = \epsilon \frac{\omega^2}{k}\sin{(kq-\omega t)}.
  \label{eq:effective_model}
\end{equation} 
The small parameter $\epsilon$ is defined as
\begin{equation}
  \epsilon = \eta\frac{A_\rho}{\rho_0},\qquad
  \mathrm{with}\quad \eta=\left(\frac{\gamma_2C_\mathrm{a}+\gamma_1}{\gamma_2C_a+\mathcal{M}}\right),
  \label{eq:epsilon}
\end{equation}
and where we have introduced the added mass coefficient ($C_\mathrm{a}=1$ in 2D) and two phenomenological dimensionless parameters $\gamma_1\simeq 0.69$ and $\gamma_2\simeq 0.25$ which account for the presence of a healing layer at the particle boundary; these values were obtained by fitting our theoretical prediction above using a small subset of simulations (see SM).

We want to establish the behaviour of the impurity position $ q $ at long times: introducing the slow time scale $ \tau=\epsilon t $ and using a standard multi-scale expansion, see Section III of the SM, we obtain the following expression for the drift velocity, i.e. the impurity velocity averaged over the fast timescale $ t $
\begin{equation}
  v_\mathrm{drift} = \left\langle\dot{q}\right\rangle_t =  - \frac{\omega}{k}\epsilon\cos(\varphi) + \frac{\omega}{k}\epsilon^2\left( 1 + \frac{1}{4}\cos{(2\varphi)} \right) + \mathcal{O}(\epsilon^3).
  \label{Eq:drift}
\end{equation}
This theoretical prediction (dashed lines of Fig.\ref{Fig:phase}) is in very good agreement with data of GP numerical simulations. 
Note that the drift velocity averaged over the impurity-wave phases vanishes at order $\mathcal{O}(\epsilon)$. 
However, the next-to-leading order remains finite:
\begin{equation}
  \left\langle v_\mathrm{drift} \right\rangle_\varphi = \frac{\omega}{k}\epsilon^2 + \mathcal{O}(\epsilon^3).
  \label{Eq:second_order}
\end{equation} 
It is interesting to notice that such result is equivalent to the Stokes drift in classical fluids for perfect tracers (see Section IV of the SM) $ v_\mathrm{drift}^\mathrm{tracer} = \frac{\omega}{2k}( {A_\rho}/{\rho_0} )^2 $, a part from a coefficient that depends in this case on $\mathcal{M}$.

\begin{figure}
  \centering
  \includegraphics[width=1\linewidth]{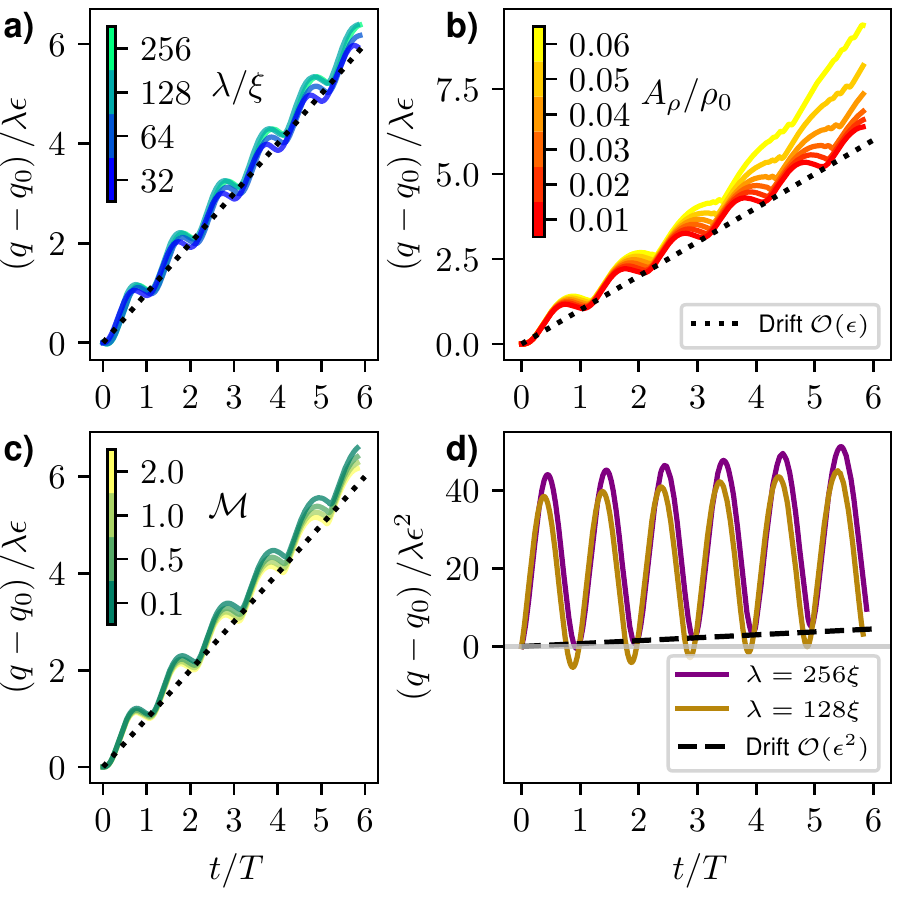}
  \caption{Time evolution of the impurity rescaled position with drift parameter $ \epsilon $ for \textbf{a)} waves of different wavelength, \textbf{b)} waves of different amplitude and \textbf{c)} impurities of different mass. Dotted lines represent the drift prediction \eqref{Eq:drift} at the leading order. \textbf{d)} Time evolution of the impurity rescaled position with drift parameter $ \epsilon^2 $ for waves of different wavelengths and same initial impurity-wave phase $\varphi=\pi/2$; the second order prediction \eqref{Eq:second_order_cheat} is displayed in dashed line.}
  \label{Fig:validation}
\end{figure}

In Fig.\ref{Fig:validation} we further validate the model \eqref{eq:effective_model} and predictions \eqref{Eq:drift} and \eqref{Eq:second_order} by varying its different parameters.
 and measuring the impurity displacement, initially set at $\varphi=\pi$, i.e, the wave trough, to highlight $ \mathcal{O}(\epsilon) $ effects. 
First we set $\varphi=\pi$, i.e the impurity at the wave trough, to highlight $ \mathcal{O}(\epsilon) $ effects: we consider waves of different wavelengths, Fig.\ref{Fig:validation}.a), and amplitudes, Fig.\ref{Fig:validation}.b), as well as impurity of different masses, Fig.\ref{Fig:validation}.c). 
In all the cases studied we observe that the motion curves of the particle collapse when the time is normalised by the wave period $T$ and the displacement by $\lambda\epsilon$.
Then, in Fig.\ref{Fig:validation}.d, we also check the prediction for the drift at the order $\mathcal{O}(\epsilon^2)$: since averaging over all the impurity-wave phases is computational demanding, we consider only the initial phase $\varphi=\pi/2$, for which the leading order vanishes. 
Equation \eqref{Eq:drift} results in
\begin{equation}
 v_\mathrm{drift}(\varphi=\pi/2) = \frac{3\omega}{4k}\epsilon^2 + \mathcal{O}(\epsilon^3),
  \label{Eq:second_order_cheat}
\end{equation} 
which fits well the numerical data.
Overall, we conclude that predictions \eqref{Eq:drift} and \eqref{Eq:second_order} are robust up to $\mathcal{O}(\epsilon^2)$.

In summary, we have described and explained how a impurity immersed in a quantum fluid experiences a net transfer of momentum from an incoming density wave due to Stokes drift. 
At leading-order, the impurity drift depends on the initial impurity position with respect to the phase of the incoming wave: remarkably, it can move in any direction, independently on the direction of the incoming wave. 
When averaging over the initial impurity position with respect to the wave phase, this first-order effects cancel out and a non-vanishing second-order drift exits along the direction of the wave; this is consistent with the classical Stokes drift, but with a different coefficient.

The Stokes drift and impurity transport theoretical predictions reported in this Letter are derived under the assumption of a classical passive impurities, and were further corroborated by GP numerical results whit active impurities. 
This constitutes evidence of their quantitative applicability to quantum fluid experiments like superfluids of light made in photorefractive crystals. 
In a recent experiment \cite{michelSuperfluidMotionDragforce2018}, the dynamics of a very small defect ($\sim 1\xi$), obtained by imprinting a depletion in one side of the crystal, was studied to address the breakdown of superfluidity. 
Although the current propagation distance of the crystal remains relatively short ($\sim 10\xi $), it should be possible to observe the first-order drift effect. In order to observe the second order correction, a larger system would be needed; such requirement is challenging in current experiments but might be overcome on the future. 

Finally, it is worth mentioning that, as manifested in Fig.~\ref{Fig:density_evo}, an interesting effect appears at long times. 
After about 13 linear periods $ T $ of interaction with the impurity (note that our system is periodic), the incoming wave deforms into a localised coherent structure that resembles a grey soliton. 
We expect the emergence of such nonlinear structures to be enhanced for larger incoming wave amplitudes. 
In this limit, the theory developed in this Letter might fail, which is consistent with the deviations observed in Fig.~\ref{Fig:validation}.b) for large amplitudes. 
The emergence and interaction of a soliton, or a train of solitons, and an impurity are interesting new lines of investigation that should be addressed in future works.

\begin{acknowledgments}
GK and MO acknowledge the support of the Simons Foundation Collaboration grant Wave Turbulence (Award ID 651471).
GK was funded by the Agence Nationale de la Recherche through the project GIANTE ANR-18-CE30-0020-01.
DP was supported by the EPSRC First Grant Number EP/P023770/1.
Computations were carried out at the M{\'e}socentre SIGAMM hosted at the Observatoire de la C{\^o}te d'Azur.
This research was originally conceived to celebrate the bicentenary of the birth of Sir George Gabriel Stokes occurred the 13th August 1819.
DP would like to thank the Isaac Newton Institute for Mathematical Sciences for support and hospitality during the programme Dispersive hydrodynamics: mathematics, simulation and experiments when the final part of this work was undertaken, supported by EPSRC Grant Number EP/R014604/1.
\end{acknowledgments}

\bibliographystyle{apsrev4-1}
\bibliography{../references}

\begin{thebibliography}{25}%
\makeatletter
\providecommand \@ifxundefined [1]{%
 \@ifx{#1\undefined}
}%
\providecommand \@ifnum [1]{%
 \ifnum #1\expandafter \@firstoftwo
 \else \expandafter \@secondoftwo
 \fi
}%
\providecommand \@ifx [1]{%
 \ifx #1\expandafter \@firstoftwo
 \else \expandafter \@secondoftwo
 \fi
}%
\providecommand \natexlab [1]{#1}%
\providecommand \enquote  [1]{``#1''}%
\providecommand \bibnamefont  [1]{#1}%
\providecommand \bibfnamefont [1]{#1}%
\providecommand \citenamefont [1]{#1}%
\providecommand \href@noop [0]{\@secondoftwo}%
\providecommand \href [0]{\begingroup \@sanitize@url \@href}%
\providecommand \@href[1]{\@@startlink{#1}\@@href}%
\providecommand \@@href[1]{\endgroup#1\@@endlink}%
\providecommand \@sanitize@url [0]{\catcode `\\12\catcode `\$12\catcode
  `\&12\catcode `\#12\catcode `\^12\catcode `\_12\catcode `\%12\relax}%
\providecommand \@@startlink[1]{}%
\providecommand \@@endlink[0]{}%
\providecommand \url  [0]{\begingroup\@sanitize@url \@url }%
\providecommand \@url [1]{\endgroup\@href {#1}{\urlprefix }}%
\providecommand \urlprefix  [0]{URL }%
\providecommand \Eprint [0]{\href }%
\providecommand \doibase [0]{http://dx.doi.org/}%
\providecommand \selectlanguage [0]{\@gobble}%
\providecommand \bibinfo  [0]{\@secondoftwo}%
\providecommand \bibfield  [0]{\@secondoftwo}%
\providecommand \translation [1]{[#1]}%
\providecommand \BibitemOpen [0]{}%
\providecommand \bibitemStop [0]{}%
\providecommand \bibitemNoStop [0]{.\EOS\space}%
\providecommand \EOS [0]{\spacefactor3000\relax}%
\providecommand \BibitemShut  [1]{\csname bibitem#1\endcsname}%
\let\auto@bib@innerbib\@empty
\bibitem [{\citenamefont {Stokes}(1880)}]{stokes1880theory}%
  \BibitemOpen
  \bibfield  {author} {\bibinfo {author} {\bibfnamefont {G.~G.}\ \bibnamefont
  {Stokes}},\ }\href@noop {} {\bibfield  {journal} {\bibinfo  {journal}
  {Transactions of the Cambridge philosophical society}\ } (\bibinfo {year}
  {1880})}\BibitemShut {NoStop}%
\bibitem [{\citenamefont {Longuet-Higgins}(1953)}]{longuet1953mass}%
  \BibitemOpen
  \bibfield  {author} {\bibinfo {author} {\bibfnamefont {M.~S.}\ \bibnamefont
  {Longuet-Higgins}},\ }\href@noop {} {\bibfield  {journal} {\bibinfo
  {journal} {Philosophical Transactions of the Royal Society of London. Series
  A, Mathematical and Physical Sciences}\ }\textbf {\bibinfo {volume} {245}},\
  \bibinfo {pages} {535} (\bibinfo {year} {1953})}\BibitemShut {NoStop}%
\bibitem [{\citenamefont {Besio}\ \emph {et~al.}(2004)\citenamefont {Besio},
  \citenamefont {Blondeaux}, \citenamefont {Brocchini},\ and\ \citenamefont
  {Vittori}}]{besio2004modeling}%
  \BibitemOpen
  \bibfield  {author} {\bibinfo {author} {\bibfnamefont {G.}~\bibnamefont
  {Besio}}, \bibinfo {author} {\bibfnamefont {P.}~\bibnamefont {Blondeaux}},
  \bibinfo {author} {\bibfnamefont {M.}~\bibnamefont {Brocchini}}, \ and\
  \bibinfo {author} {\bibfnamefont {G.}~\bibnamefont {Vittori}},\ }\href@noop
  {} {\bibfield  {journal} {\bibinfo  {journal} {Journal of Geophysical
  Research: Oceans}\ }\textbf {\bibinfo {volume} {109}} (\bibinfo {year}
  {2004})}\BibitemShut {NoStop}%
\bibitem [{\citenamefont {Santamaria}\ \emph {et~al.}(2013)\citenamefont
  {Santamaria}, \citenamefont {Boffetta}, \citenamefont {Afonso}, \citenamefont
  {Mazzino}, \citenamefont {Onorato},\ and\ \citenamefont
  {Pugliese}}]{santamaria2013stokes}%
  \BibitemOpen
  \bibfield  {author} {\bibinfo {author} {\bibfnamefont {F.}~\bibnamefont
  {Santamaria}}, \bibinfo {author} {\bibfnamefont {G.}~\bibnamefont
  {Boffetta}}, \bibinfo {author} {\bibfnamefont {M.~M.}\ \bibnamefont
  {Afonso}}, \bibinfo {author} {\bibfnamefont {A.}~\bibnamefont {Mazzino}},
  \bibinfo {author} {\bibfnamefont {M.}~\bibnamefont {Onorato}}, \ and\
  \bibinfo {author} {\bibfnamefont {D.}~\bibnamefont {Pugliese}},\ }\href@noop
  {} {\bibfield  {journal} {\bibinfo  {journal} {EPL (Europhysics Letters)}\
  }\textbf {\bibinfo {volume} {102}},\ \bibinfo {pages} {14003} (\bibinfo
  {year} {2013})}\BibitemShut {NoStop}%
\bibitem [{\citenamefont {Bruus}(2012)}]{bruus2012acoustofluidics}%
  \BibitemOpen
  \bibfield  {author} {\bibinfo {author} {\bibfnamefont {H.}~\bibnamefont
  {Bruus}},\ }\href@noop {} {\bibfield  {journal} {\bibinfo  {journal} {Lab on
  a Chip}\ }\textbf {\bibinfo {volume} {12}},\ \bibinfo {pages} {1014}
  (\bibinfo {year} {2012})}\BibitemShut {NoStop}%
\bibitem [{\citenamefont {Toftul}\ \emph {et~al.}(2019)\citenamefont {Toftul},
  \citenamefont {Bliokh}, \citenamefont {Petrov},\ and\ \citenamefont
  {Nori}}]{toftul2019acoustic}%
  \BibitemOpen
  \bibfield  {author} {\bibinfo {author} {\bibfnamefont {I.}~\bibnamefont
  {Toftul}}, \bibinfo {author} {\bibfnamefont {K.}~\bibnamefont {Bliokh}},
  \bibinfo {author} {\bibfnamefont {M.~I.}\ \bibnamefont {Petrov}}, \ and\
  \bibinfo {author} {\bibfnamefont {F.}~\bibnamefont {Nori}},\ }\href@noop {}
  {\bibfield  {journal} {\bibinfo  {journal} {Physical review letters}\
  }\textbf {\bibinfo {volume} {123}},\ \bibinfo {pages} {183901} (\bibinfo
  {year} {2019})}\BibitemShut {NoStop}%
\bibitem [{\citenamefont {Xie}\ and\ \citenamefont
  {Vanneste}(2014)}]{xie2014dynamics}%
  \BibitemOpen
  \bibfield  {author} {\bibinfo {author} {\bibfnamefont {J.-H.}\ \bibnamefont
  {Xie}}\ and\ \bibinfo {author} {\bibfnamefont {J.}~\bibnamefont {Vanneste}},\
  }\href@noop {} {\bibfield  {journal} {\bibinfo  {journal} {Physics of
  Fluids}\ }\textbf {\bibinfo {volume} {26}},\ \bibinfo {pages} {102001}
  (\bibinfo {year} {2014})}\BibitemShut {NoStop}%
\bibitem [{\citenamefont {Cleckler}\ \emph {et~al.}(2012)\citenamefont
  {Cleckler}, \citenamefont {Elghobashi},\ and\ \citenamefont
  {Liu}}]{cleckler2012motion}%
  \BibitemOpen
  \bibfield  {author} {\bibinfo {author} {\bibfnamefont {J.}~\bibnamefont
  {Cleckler}}, \bibinfo {author} {\bibfnamefont {S.}~\bibnamefont
  {Elghobashi}}, \ and\ \bibinfo {author} {\bibfnamefont {F.}~\bibnamefont
  {Liu}},\ }\href@noop {} {\bibfield  {journal} {\bibinfo  {journal} {Physics
  of Fluids}\ }\textbf {\bibinfo {volume} {24}},\ \bibinfo {pages} {033301}
  (\bibinfo {year} {2012})}\BibitemShut {NoStop}%
\bibitem [{\citenamefont {King}(1934)}]{king1934acoustic}%
  \BibitemOpen
  \bibfield  {author} {\bibinfo {author} {\bibfnamefont {L.~V.}\ \bibnamefont
  {King}},\ }\href@noop {} {\bibfield  {journal} {\bibinfo  {journal}
  {Proceedings of the Royal Society of London. Series A-Mathematical and
  Physical Sciences}\ }\textbf {\bibinfo {volume} {147}},\ \bibinfo {pages}
  {212} (\bibinfo {year} {1934})}\BibitemShut {NoStop}%
\bibitem [{\citenamefont {Poole}\ \emph {et~al.}(2005)\citenamefont {Poole},
  \citenamefont {Barenghi}, \citenamefont {Sergeev},\ and\ \citenamefont
  {Vinen}}]{PhysRevB.71.064514}%
  \BibitemOpen
  \bibfield  {author} {\bibinfo {author} {\bibfnamefont {D.~R.}\ \bibnamefont
  {Poole}}, \bibinfo {author} {\bibfnamefont {C.~F.}\ \bibnamefont {Barenghi}},
  \bibinfo {author} {\bibfnamefont {Y.~A.}\ \bibnamefont {Sergeev}}, \ and\
  \bibinfo {author} {\bibfnamefont {W.~F.}\ \bibnamefont {Vinen}},\ }\href
  {\doibase 10.1103/PhysRevB.71.064514} {\bibfield  {journal} {\bibinfo
  {journal} {Phys. Rev. B}\ }\textbf {\bibinfo {volume} {71}},\ \bibinfo
  {pages} {064514} (\bibinfo {year} {2005})}\BibitemShut {NoStop}%
\bibitem [{\citenamefont {Polanco}\ and\ \citenamefont
  {Krstulovic}(2020)}]{PhysRevFluids.5.032601}%
  \BibitemOpen
  \bibfield  {author} {\bibinfo {author} {\bibfnamefont {J.~I.}\ \bibnamefont
  {Polanco}}\ and\ \bibinfo {author} {\bibfnamefont {G.}~\bibnamefont
  {Krstulovic}},\ }\href {\doibase 10.1103/PhysRevFluids.5.032601} {\bibfield
  {journal} {\bibinfo  {journal} {Phys. Rev. Fluids}\ }\textbf {\bibinfo
  {volume} {5}},\ \bibinfo {pages} {032601} (\bibinfo {year}
  {2020})}\BibitemShut {NoStop}%
\bibitem [{\citenamefont {Sergeev}\ and\ \citenamefont
  {Barenghi}(2009)}]{Sergeev:2009un}%
  \BibitemOpen
  \bibfield  {author} {\bibinfo {author} {\bibfnamefont {Y.~A.}\ \bibnamefont
  {Sergeev}}\ and\ \bibinfo {author} {\bibfnamefont {C.~F.}\ \bibnamefont
  {Barenghi}},\ }\href {\doibase 10.1007/s10909-009-9994-8} {\bibfield
  {journal} {\bibinfo  {journal} {Journal of Low Temperature Physics}\ }\textbf
  {\bibinfo {volume} {157}},\ \bibinfo {pages} {429} (\bibinfo {year}
  {2009})}\BibitemShut {NoStop}%
\bibitem [{\citenamefont {Giuriato}\ and\ \citenamefont
  {Krstulovic}(2019)}]{giuriatoInteractionActiveParticles2019}%
  \BibitemOpen
  \bibfield  {author} {\bibinfo {author} {\bibfnamefont {U.}~\bibnamefont
  {Giuriato}}\ and\ \bibinfo {author} {\bibfnamefont {G.}~\bibnamefont
  {Krstulovic}},\ }\href {\doibase 10.1038/s41598-019-39877-w} {\bibfield
  {journal} {\bibinfo  {journal} {Scientific Reports}\ }\textbf {\bibinfo
  {volume} {9}},\ \bibinfo {pages} {4839} (\bibinfo {year} {2019})}\BibitemShut
  {NoStop}%
\bibitem [{\citenamefont {Williams}\ and\ \citenamefont
  {Packard}(1974)}]{PhysRevLett.33.280}%
  \BibitemOpen
  \bibfield  {author} {\bibinfo {author} {\bibfnamefont {G.~A.}\ \bibnamefont
  {Williams}}\ and\ \bibinfo {author} {\bibfnamefont {R.~E.}\ \bibnamefont
  {Packard}},\ }\href {\doibase 10.1103/PhysRevLett.33.280} {\bibfield
  {journal} {\bibinfo  {journal} {Phys. Rev. Lett.}\ }\textbf {\bibinfo
  {volume} {33}},\ \bibinfo {pages} {280} (\bibinfo {year} {1974})}\BibitemShut
  {NoStop}%
\bibitem [{\citenamefont {Bewley}\ \emph {et~al.}(2008)\citenamefont {Bewley},
  \citenamefont {Paoletti}, \citenamefont {Sreenivasan},\ and\ \citenamefont
  {Lathrop}}]{P.:2008tp}%
  \BibitemOpen
  \bibfield  {author} {\bibinfo {author} {\bibfnamefont {G.~P.}\ \bibnamefont
  {Bewley}}, \bibinfo {author} {\bibfnamefont {M.~S.}\ \bibnamefont
  {Paoletti}}, \bibinfo {author} {\bibfnamefont {K.~R.}\ \bibnamefont
  {Sreenivasan}}, \ and\ \bibinfo {author} {\bibfnamefont {D.~P.}\ \bibnamefont
  {Lathrop}},\ }\href {\doibase 10.1073/pnas.0806002105} {\bibfield  {journal}
  {\bibinfo  {journal} {Proceedings of the National Academy of Sciences}\
  }\textbf {\bibinfo {volume} {105}},\ \bibinfo {pages} {13707} (\bibinfo
  {year} {2008})}\BibitemShut {NoStop}%
\bibitem [{\citenamefont {{Winiecki, T.}}\ and\ \citenamefont {{Adams, C.
  S.}}(2000)}]{refId0}%
  \BibitemOpen
  \bibfield  {author} {\bibinfo {author} {\bibnamefont {{Winiecki, T.}}}\ and\
  \bibinfo {author} {\bibnamefont {{Adams, C. S.}}},\ }\href {\doibase
  10.1209/epl/i2000-00432-x} {\bibfield  {journal} {\bibinfo  {journal}
  {Europhys. Lett.}\ }\textbf {\bibinfo {volume} {52}},\ \bibinfo {pages} {257}
  (\bibinfo {year} {2000})}\BibitemShut {NoStop}%
\bibitem [{\citenamefont {Clark}(1965)}]{Clark:1965wp}%
  \BibitemOpen
  \bibfield  {author} {\bibinfo {author} {\bibfnamefont {R.~C.}\ \bibnamefont
  {Clark}},\ }\href {\doibase https://doi.org/10.1016/0031-9163(65)90395-1}
  {\bibfield  {journal} {\bibinfo  {journal} {Physics Letters}\ }\textbf
  {\bibinfo {volume} {16}},\ \bibinfo {pages} {42} (\bibinfo {year}
  {1965})}\BibitemShut {NoStop}%
\bibitem [{\citenamefont {Berloff}\ and\ \citenamefont
  {Roberts}(2000)}]{PhysRevB.63.024510}%
  \BibitemOpen
  \bibfield  {author} {\bibinfo {author} {\bibfnamefont {N.~G.}\ \bibnamefont
  {Berloff}}\ and\ \bibinfo {author} {\bibfnamefont {P.~H.}\ \bibnamefont
  {Roberts}},\ }\href {\doibase 10.1103/PhysRevB.63.024510} {\bibfield
  {journal} {\bibinfo  {journal} {Phys. Rev. B}\ }\textbf {\bibinfo {volume}
  {63}},\ \bibinfo {pages} {024510} (\bibinfo {year} {2000})}\BibitemShut
  {NoStop}%
\bibitem [{\citenamefont {Villois}\ and\ \citenamefont
  {Salman}(2018)}]{PhysRevB.97.094507}%
  \BibitemOpen
  \bibfield  {author} {\bibinfo {author} {\bibfnamefont {A.}~\bibnamefont
  {Villois}}\ and\ \bibinfo {author} {\bibfnamefont {H.}~\bibnamefont
  {Salman}},\ }\href {\doibase 10.1103/PhysRevB.97.094507} {\bibfield
  {journal} {\bibinfo  {journal} {Phys. Rev. B}\ }\textbf {\bibinfo {volume}
  {97}},\ \bibinfo {pages} {094507} (\bibinfo {year} {2018})}\BibitemShut
  {NoStop}%
\bibitem [{\citenamefont {Giuriato}\ \emph {et~al.}(2020)\citenamefont
  {Giuriato}, \citenamefont {Krstulovic},\ and\ \citenamefont
  {Nazarenko}}]{giuriatoHowTrappedParticles2020}%
  \BibitemOpen
  \bibfield  {author} {\bibinfo {author} {\bibfnamefont {U.}~\bibnamefont
  {Giuriato}}, \bibinfo {author} {\bibfnamefont {G.}~\bibnamefont
  {Krstulovic}}, \ and\ \bibinfo {author} {\bibfnamefont {S.}~\bibnamefont
  {Nazarenko}},\ }\href {\doibase 10.1103/PhysRevResearch.2.023149} {\bibfield
  {journal} {\bibinfo  {journal} {Physical Review Research}\ }\textbf {\bibinfo
  {volume} {2}},\ \bibinfo {pages} {023149} (\bibinfo {year}
  {2020})}\BibitemShut {NoStop}%
\bibitem [{\citenamefont {Giuriato}\ and\ \citenamefont
  {Krstulovic}(2021)}]{giuriatoStochasticMotionFinitesize2021}%
  \BibitemOpen
  \bibfield  {author} {\bibinfo {author} {\bibfnamefont {U.}~\bibnamefont
  {Giuriato}}\ and\ \bibinfo {author} {\bibfnamefont {G.}~\bibnamefont
  {Krstulovic}},\ }\href {\doibase 10.1103/PhysRevB.103.024509} {\bibfield
  {journal} {\bibinfo  {journal} {Physical Review B}\ }\textbf {\bibinfo
  {volume} {103}},\ \bibinfo {pages} {024509} (\bibinfo {year}
  {2021})}\BibitemShut {NoStop}%
\bibitem [{\citenamefont {Shukla}\ \emph {et~al.}(2016)\citenamefont {Shukla},
  \citenamefont {Brachet},\ and\ \citenamefont
  {Pandit}}]{shuklaStickingTransitionMinimal2016}%
  \BibitemOpen
  \bibfield  {author} {\bibinfo {author} {\bibfnamefont {V.}~\bibnamefont
  {Shukla}}, \bibinfo {author} {\bibfnamefont {M.}~\bibnamefont {Brachet}}, \
  and\ \bibinfo {author} {\bibfnamefont {R.}~\bibnamefont {Pandit}},\ }\href
  {\doibase 10.1103/PhysRevA.94.041602} {\bibfield  {journal} {\bibinfo
  {journal} {Physical Review A}\ }\textbf {\bibinfo {volume} {94}},\ \bibinfo
  {pages} {041602} (\bibinfo {year} {2016})}\BibitemShut {NoStop}%
\bibitem [{\citenamefont {Giuriato}\ and\ \citenamefont
  {Krstulovic}(2020)}]{giuriatoQuantumVortexReconnections2020}%
  \BibitemOpen
  \bibfield  {author} {\bibinfo {author} {\bibfnamefont {U.}~\bibnamefont
  {Giuriato}}\ and\ \bibinfo {author} {\bibfnamefont {G.}~\bibnamefont
  {Krstulovic}},\ }\href {\doibase 10.1103/PhysRevB.102.094508} {\bibfield
  {journal} {\bibinfo  {journal} {Physical Review B}\ }\textbf {\bibinfo
  {volume} {102}},\ \bibinfo {pages} {094508} (\bibinfo {year}
  {2020})}\BibitemShut {NoStop}%
\bibitem [{\citenamefont {Krstulovic}(2020)}]{krstulovicHDR}%
  \BibitemOpen
  \bibfield  {author} {\bibinfo {author} {\bibfnamefont {G.}~\bibnamefont
  {Krstulovic}},\ }\emph {\bibinfo {title} {{A theoretical description of
  vortex dynamics in superfluids. Kelvin waves, reconnections and
  particle-vortex interaction}}},\ \href
  {https://hal.archives-ouvertes.fr/tel-03544830} {\bibinfo {type}
  {Habilitation {\`a} diriger des recherches}},\ \bibinfo  {school}
  {{Universit{\'e} C{\^o}te d'Azur}} (\bibinfo {year} {2020})\BibitemShut
  {NoStop}%
\bibitem [{\citenamefont {Michel}\ \emph {et~al.}(2018)\citenamefont {Michel},
  \citenamefont {Boughdad}, \citenamefont {Albert}, \citenamefont {Larre},\
  and\ \citenamefont {Bellec}}]{michelSuperfluidMotionDragforce2018}%
  \BibitemOpen
  \bibfield  {author} {\bibinfo {author} {\bibfnamefont {C.}~\bibnamefont
  {Michel}}, \bibinfo {author} {\bibfnamefont {O.}~\bibnamefont {Boughdad}},
  \bibinfo {author} {\bibfnamefont {M.}~\bibnamefont {Albert}}, \bibinfo
  {author} {\bibfnamefont {P.-E.}\ \bibnamefont {Larre}}, \ and\ \bibinfo
  {author} {\bibfnamefont {M.}~\bibnamefont {Bellec}},\ }\href {\doibase
  10.1038/s41467-018-04534-9} {\bibfield  {journal} {\bibinfo  {journal}
  {Nature Communications}\ }\textbf {\bibinfo {volume} {9}},\ \bibinfo {pages}
  {2108} (\bibinfo {year} {2018})}\BibitemShut {NoStop}%
\end{thebibliography}%


\begin{thebibliography}{3}%
\makeatletter
\providecommand \@ifxundefined [1]{%
 \@ifx{#1\undefined}
}%
\providecommand \@ifnum [1]{%
 \ifnum #1\expandafter \@firstoftwo
 \else \expandafter \@secondoftwo
 \fi
}%
\providecommand \@ifx [1]{%
 \ifx #1\expandafter \@firstoftwo
 \else \expandafter \@secondoftwo
 \fi
}%
\providecommand \natexlab [1]{#1}%
\providecommand \enquote  [1]{``#1''}%
\providecommand \bibnamefont  [1]{#1}%
\providecommand \bibfnamefont [1]{#1}%
\providecommand \citenamefont [1]{#1}%
\providecommand \href@noop [0]{\@secondoftwo}%
\providecommand \href [0]{\begingroup \@sanitize@url \@href}%
\providecommand \@href[1]{\@@startlink{#1}\@@href}%
\providecommand \@@href[1]{\endgroup#1\@@endlink}%
\providecommand \@sanitize@url [0]{\catcode `\\12\catcode `\$12\catcode
  `\&12\catcode `\#12\catcode `\^12\catcode `\_12\catcode `\%12\relax}%
\providecommand \@@startlink[1]{}%
\providecommand \@@endlink[0]{}%
\providecommand \url  [0]{\begingroup\@sanitize@url \@url }%
\providecommand \@url [1]{\endgroup\@href {#1}{\urlprefix }}%
\providecommand \urlprefix  [0]{URL }%
\providecommand \Eprint [0]{\href }%
\providecommand \doibase [0]{http://dx.doi.org/}%
\providecommand \selectlanguage [0]{\@gobble}%
\providecommand \bibinfo  [0]{\@secondoftwo}%
\providecommand \bibfield  [0]{\@secondoftwo}%
\providecommand \translation [1]{[#1]}%
\providecommand \BibitemOpen [0]{}%
\providecommand \bibitemStop [0]{}%
\providecommand \bibitemNoStop [0]{.\EOS\space}%
\providecommand \EOS [0]{\spacefactor3000\relax}%
\providecommand \BibitemShut  [1]{\csname bibitem#1\endcsname}%
\let\auto@bib@innerbib\@empty
\bibitem [{\citenamefont {Shukla}\ \emph {et~al.}(2016)\citenamefont {Shukla},
  \citenamefont {Brachet},\ and\ \citenamefont
  {Pandit}}]{shuklaStickingTransitionMinimal2016}%
  \BibitemOpen
  \bibfield  {author} {\bibinfo {author} {\bibfnamefont {V.}~\bibnamefont
  {Shukla}}, \bibinfo {author} {\bibfnamefont {M.}~\bibnamefont {Brachet}}, \
  and\ \bibinfo {author} {\bibfnamefont {R.}~\bibnamefont {Pandit}},\ }\href
  {\doibase 10.1103/PhysRevA.94.041602} {\bibfield  {journal} {\bibinfo
  {journal} {Physical Review A}\ }\textbf {\bibinfo {volume} {94}},\ \bibinfo
  {pages} {041602} (\bibinfo {year} {2016})}\BibitemShut {NoStop}%
\bibitem [{\citenamefont {Giuriato}\ and\ \citenamefont
  {Krstulovic}(2019)}]{giuriatoInteractionActiveParticles2019}%
  \BibitemOpen
  \bibfield  {author} {\bibinfo {author} {\bibfnamefont {U.}~\bibnamefont
  {Giuriato}}\ and\ \bibinfo {author} {\bibfnamefont {G.}~\bibnamefont
  {Krstulovic}},\ }\href {\doibase 10.1038/s41598-019-39877-w} {\bibfield
  {journal} {\bibinfo  {journal} {Scientific Reports}\ }\textbf {\bibinfo
  {volume} {9}},\ \bibinfo {pages} {4839} (\bibinfo {year} {2019})}\BibitemShut
  {NoStop}%
\bibitem [{\citenamefont {Maxey}(1983)}]{maxeyEquationMotionSmall1983}%
  \BibitemOpen
  \bibfield  {author} {\bibinfo {author} {\bibfnamefont {M.~R.}\ \bibnamefont
  {Maxey}},\ }\href {\doibase 10.1063/1.864230} {\bibfield  {journal} {\bibinfo
   {journal} {Physics of Fluids}\ }\textbf {\bibinfo {volume} {26}},\ \bibinfo
  {pages} {883} (\bibinfo {year} {1983})}\BibitemShut {NoStop}%
\end{thebibliography}%

%

\end{document}


\title{SUPPLEMENTAL MATERIAL: Stokes drift and impurity transport in a quantum fluid}
\author{Umberto Giuriato}
\affiliation{
Universit\'e C\^ote d'Azur, Observatoire de la C\^ote d'Azur, CNRS, Laboratoire Lagrange, Bd de l'Observatoire, CS 34229, 06304 Nice cedex 4, France.}
\author{Giorgio Krstulovic}
\affiliation{
Universit\'e C\^ote d'Azur, Observatoire de la C\^ote d'Azur, CNRS, Laboratoire Lagrange, Bd de l'Observatoire, CS 34229, 06304 Nice cedex 4, France.}
\author{Miguel Onorato}
\affiliation{Dipartimento di Fisica Generale, Universit\`a degli Studi di Torino, Via Pietro Giuria 1, 10126 Torino, Italy}
\author{Davide Proment}
\affiliation{School of Mathematics, University of East Anglia, Norwich Research Park, Norwich, NR4 7TJ, United Kingdom}

\begin{abstract}

\end{abstract}
\maketitle

\section{Quantum fluid density waves}
We consider  the Hamiltonian of a quantum fluid with an active classical impurity of mass $M_\mathrm{p}$ immersed in it \cite{shuklaStickingTransitionMinimal2016,giuriatoInteractionActiveParticles2019}:
\begin{equation}
H=\frac{{\mathbf{p}}^2}{2 \Mp}+\int\left( \frac{\hbar^2}{2m} |\grad \psi |^2 +\frac{g}{2}|\psi|^4 - \mu|\psi|^2 + V_\mathrm{p}(| \mathbf{x} -{\mathbf{ q}} |)|\psi|^2 \right) \mathrm{d} \mathbf{x} ,
\label{Eq:H}
\end{equation}
where $\mathbf{q}$  and $\mathbf{p}$ are the impurity position and momentum, while $V_\mathrm{p}$ is the repulsive potential localized around $\mathbf{q}$ which models the impurity shape.
In absence of impurities, the Hamiltonian models the Gross-Pitaevskii equation that reads:
\begin{equation}
i\hbar\frac{\partial\psi}{\partial t} = -\frac{\hbar^2}{2m}\nabla^2\psi + g|\psi|^2\psi - \mu\psi.
\label{Eq:GP}
\end{equation}
Here the uniform ground state is fixed by the chemical potential $\psi_0=\sqrt{\mu/g}$. Given the Madelung transformation $\psi(\mathbf{x},t)=\sqrt{\frac{\rho(\mathbf{x},t)}{m}}e^{i\frac{m}{\hbar}\phi(\mathbf{x},t)}$, the corresponding hydrodynamic equations are
\begin{eqnarray}
\frac{\partial \rho}{\partial t} + \nabla\cdot\rho\mathbf{v} = 0\\
\frac{\partial \phi}{\partial t} + \frac{1}{2}|\nabla\phi|^2 + \frac{g}{m^2}(\rho-\rho_0) - \frac{\hbar^2}{2m^2}\frac{\nabla^2\sqrt{\rho}}{\sqrt{\rho}}=0
\label{Eq:hydro}
\end{eqnarray}
where $\mathbf{v}=\nabla\phi$. The pressure term is composed by a classical and a quantum term $p=p_\mathrm{cl}+p_\mathrm{q}$:
\begin{equation}
\frac{p_\mathrm{cl}}{\rho} =  \frac{g}{m^2}(\rho-\rho_0),\qquad\quad\frac{p_\mathrm{q}}{\rho} =  - \frac{\hbar^2}{2m^2}\frac{\nabla^2\sqrt{\rho}}{\sqrt{\rho}}.
\label{Eq:pressure}
\end{equation}

We consider a slightly perturbed ground state density $\psi_\mathrm{w}(\mathbf{x},t)=\sqrt{\frac{\rho_0+\rho_\mathrm{w}(\mathbf{x},t)}{m}}e^{i\frac{m}{\hbar}\phi_\mathrm{w}(\mathbf{x},t)}$, where $ |\rho_\mathrm{w}(\mathbf{x},t)|\ll \rho_0$. We consider a 1D perturbation along the $x$-axis. Linearizing around the ground state $\rho_0$ and separating real and imaginary parts, we get the linearized hydrodynamic equations:
\begin{eqnarray}
\frac{\partial \rho_\mathrm{w}}{\partial t} + \rho_0\frac{\partial v_\mathrm{w}}{\partial x} = 0.  \label{Eq:rho}\\
\frac{\partial v_\mathrm{w}}{\partial t} + \frac{g}{m^2}\frac{\partial\rho_\mathrm{w}}{\partial x} - \frac{\hbar^2}{4m^2\rho_0}\frac{\partial^3\rho_\mathrm{w}}{\partial x^3}= 0
\label{Eq:v}
\end{eqnarray}
where $v_\mathrm{w} = \partial \phi/\partial x$.
The system (\ref{Eq:rho}-\ref{Eq:v}) admits a traveling density wave solution with density amplitude $A_\rho$, provided that $A_\rho\ll\rho_0$:
\begin{equation}
\rho = \rho_0 + A_\rho\cos{(kx-\omega t)},\qquad\quad v_\mathrm{w} = \frac{\omega A_\rho}{k\rho_0}\cos{(kx-\omega t)}
\label{Eq:wave}
\end{equation}
with the dispersion relation 
\begin{equation}
\omega = c|k|\sqrt{1+\frac{\xi^2k^2}{2}}
\label{Eq:bogo}
\end{equation}
where $c=\sqrt{g\rho_0/m^2}$ is the speed of sound and $\xi=\hbar/\sqrt{2g\rho_0}$ is the healing length.

\clearpage
\section{Effective model for impurity-wave interaction}
\begin{figure}
   \centering
  \includegraphics[width=.5\linewidth]{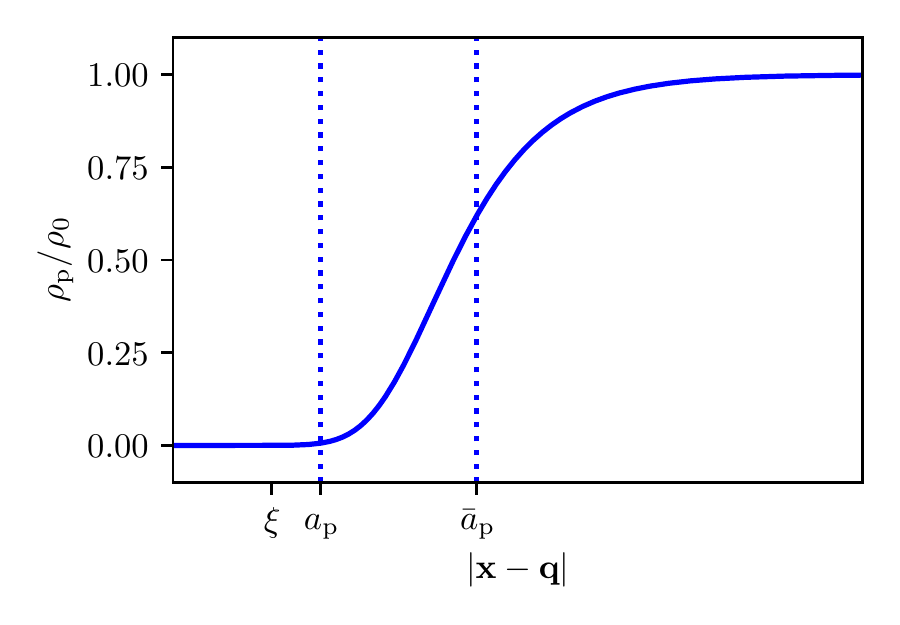}
  \caption{Quantum fluid density depletion generated by a small impurity. The vertical dotted lines indicate the hard-core size $a_\mathrm{p}$ and the effective size $\bar{a}_\mathrm{p}$.}
  \label{Fig:part}
\end{figure}

When the impurity is present, the ground state $\psi_\mathrm{p}$ corresponds to a uniform condensate far from the impurity and a density depletion at places where $|\mathbf{x}-\mathbf{q}|<a_\mathrm{p}$, with $a_\mathrm{p}$ the distance at which $\Vp(a_\mathrm{p})=\mu$. At the boundary of the impurity the density goes from zero to $\rho_0$ in a distance of the order of the healing length $\xi$. The mass of the displaced quantum fluid is defined as 
\begin{equation}
    M_0 = \rho_0\int (|\psi_0|^2 - |\psi_\mathrm{p}|^2)\,\mathrm{d}\mathbf{x}, 
\label{Eq:Mass}
\end{equation}
from which we can define the effective size of the impurity as 
$\bar{a}_\mathrm{p}=\sqrt{M_0/\pi\rho_0}$ (in 2D, assuming polar symmetry).
Figure~\ref{Fig:part} displays the quantum fluid density depletion generated by an impurity, the corresponding hard-core size $a_\mathrm{p}$ and effective size $\bar{a}_\mathrm{p}$.
A good approximation when $a_\mathrm{p}\gg\xi$ is given by the Thomas-Fermi ground state that is obtained neglecting the kinetic term in the GP model. 
It reads
\begin{equation}
|\psi_\mathrm{p}(\mathbf{x})|^2=\frac{\rho_{\rm p}(|\x-\q|)}{m} = \frac{\rho_0}{m} \left(1-\frac{\Vp(|\mathbf{x}-\mathbf{q}|)}{\mu}\right)\theta_\mathrm{H}\left[1-\frac{\Vp(|\mathbf{x}-\mathbf{q}|)}{\mu}\right], 
\label{Eq:RhoPart}
\end{equation}
with $\theta_\mathrm{H}$ the Heaviside function. If one neglects the healing layer, the impurity can be considered as a homogeneous disc, whose main effect is to exclude a ball $\mathbb{B}(\mathbf{q},\bar{a}_\mathrm{p})$ centered at the impurity position with radius $\bar{a}_\mathrm{p}$ from the quantum fluid domain.
In first approximation, given a density wave perturbation $\rho_\mathrm{w}$ the presence of the impurity modifies the density as $\rho=\rho_\mathrm{p}(\rho_0+\rho_\mathrm{w})/\rho_0$.

By variating the Hamiltonian (\ref{Eq:H}) with respect to $\mathbf{q}$ we obtain the equation of motion for the impurity
\begin{equation}
M_\mathrm{p}\ddot{\q}=\mathbf{F}=-\int V_\mathrm{p}(|\x-\q|)\nabla|\psi|^2\,\mathrm{d}\mathbf{x},
\label{Eq:Part}
\end{equation}
which can be rearranged as
\begin{equation}
\mathbf{F} = \frac{\mu}{m}\int\left(1- \frac{V_\mathrm{p}}{\mu} \right)\nabla\rho\,\mathrm{d}\mathbf{x} \sim 
\frac{\mu}{m}\int\left(1- \frac{V_\mathrm{p}}{\mu} \right)\frac{\rho_\mathrm{p}}{\rho_0}\nabla\rho_\mathrm{w}\,\mathrm{d}\mathbf{x}\sim
\frac{\mu}{m}\int\frac{\rho_\mathrm{p}}{\rho_0}\nabla\rho_\mathrm{w}\,\mathrm{d}\mathbf{x},
\label{Eq:Part2}
\end{equation}
where we used that $\left(1- \frac{V_\mathrm{p}}{\mu} \right)\frac{\rho_\mathrm{p}}{\rho_0}\sim\frac{\rho_\mathrm{p}}{\rho_0}$. 
Last expression can be further split into
\begin{equation}
\mathbf{F} \sim 
\frac{\mu}{m}\int\nabla\rho_\mathrm{w}\,\mathrm{d}\mathbf{x} - 
\frac{\mu}{m}\int\left( 1-\frac{\rho_\mathrm{p}}{\rho_0} \right)\nabla\rho_\mathrm{w}\,\mathrm{d}\mathbf{x},
\label{Eq:Part3}
\end{equation}
where the first integral vanishes assuming periodic boundary conditions. Finally we recognize the gradient of the pressure: 
\begin{equation}
\mathbf{F} \sim
- \rho_0\int\left( 1-\frac{\rho_\mathrm{p}}{\rho_0} \right)\nabla\left(\frac{p}{\rho}\right)\,\mathrm{d}\mathbf{x},
\label{Eq:Part3}
\end{equation}
where we assumed $p\sim p_\mathrm{cl}$. Now we can apply the Bernoulli equation $-p/\rho = \frac{\partial \phi}{\partial t} + \frac{1}{2}|\nabla\phi|^2$, where the potential flow $\phi = \phi_\mathrm{w} + \phi_\mathrm{p} + \phi_\mathrm{BC} $ takes into account the flow generated by the wave $\phi_\mathrm{w}$ and the flow due to the presence of the impurity $\phi_\mathrm{p}$:
\begin{eqnarray}
\phi_\mathrm{p}(\x;\q,\dot{\q})=-\frac{\bar{a}_\mathrm{p}^3}{2|\mathbf{x-\q}|^3}(\mathbf{x-\q})\cdot(\mathbf{\dot{q}}-\mathbf{v}_{\rm w}(\mathbf{q}))\quad\mathrm{in\,3D} ,\\
\phi_\mathrm{p}(\x;\q,\dot{\q})=-\frac{\bar{a}_\mathrm{p}^2}{|\mathbf{x-\q}|^2}(\mathbf{x-\q})\cdot(\mathbf{\dot{q}}-\mathbf{v}_{\rm w}(\mathbf{q}))\quad\mathrm{in\,2D} .
\label{Eq:phase_p}
\end{eqnarray}
The potential $\phi_\mathrm{BC}$ is in principle determined by the condition at the impurity boundary $\nabla\phi_\mathrm{BC}\cdot\mathbf{n}=\left[\mathbf{v}_{\rm w}(\mathbf{q}+\bar{a}_\mathrm{p}\mathbf{n})-\mathbf{v}_{\rm w}(\mathbf{q})\right]\cdot\mathbf{n}$. It can be neglected for small impurity ($ka_\mathrm{p}\ll 1$).

The equation of motion for a small impurity can be derived, using the standard procedure of impurity dynamics in classical fluids \cite{maxeyEquationMotionSmall1983}. We treat separately the two contributions of the flow to the force $\mathbf{F} = \mathbf{F}_\mathrm{w} + \mathbf{F}_\mathrm{p}$. The force generated by the wave flow is:
\begin{eqnarray}
\mathbf{F}_\mathrm{w} \sim - \rho_0\int \left( 1-\frac{\rho_\mathrm{p}}{\rho_0} \right) \nabla\left(\frac{\partial\phi_\mathrm{w}}{\partial t}\right)\,\mathrm{d}\mathbf{x}\sim \nonumber 
- \rho_0\int_{\mathbb{B}(\mathbf{q},a_\mathrm{p})} \frac{\partial v_\mathrm{w}}{\partial t} \,\mathrm{d}\mathbf{x} =
\gamma_1 M_0\frac{\partial v_\mathrm{w}}{\partial t}\bigg\rvert_\mathbf{q}
\label{Eq:Fw}
\end{eqnarray}
where we introduced the phenomenological parameter $\gamma_1$ to account for the compressible effects due to the healing layer.  The force generated by the flow $\phi_\mathrm{p}$ is computed using the Stokes theorem, assuming that the effective impurity boundary is placed at a distance $\bar{a}_\mathrm{p}$ from the center. In 2D:
\begin{eqnarray}
\mathbf{F}_\mathrm{p} \approx - \rho_0\int \left( 1-\frac{\rho_\mathrm{p}}{\rho_0} \right) \nabla\left(\frac{\partial\phi_\mathrm{p}}{\partial t} + \frac{1}{2}|\nabla\phi_\mathrm{p}|^2 \right)\,\mathrm{d}\mathbf{x}\approx
- \rho_0\int_{\mathbb{B}(\mathbf{q},\bar{a}_\mathrm{p})} \nabla\left(\frac{\partial\phi_\mathrm{p}}{\partial t} + \frac{1}{2}|\nabla\phi_\mathrm{p}|^2 \right)\,\mathrm{d}\mathbf{x} \nonumber \\ \approx
\rho_0\int_0^{2\pi}\left(\frac{\partial\phi_\mathrm{p}}{\partial t} + \frac{1}{2}|\nabla\phi_\mathrm{p}|^2 \right)\bigg\rvert_{|\x-\q|=\bar{a}_\mathrm{p}} \bar{a}_\mathrm{p}\cos{(\theta)}\,\mathrm{d}\theta= 
\gamma_2 C_\mathrm{a}M_0\left( \frac{\partial v_\mathrm{w}}{\partial t}\bigg\rvert_\mathbf{q} - \ddot{\q} \right),
\label{Eq:Fp}
\end{eqnarray}
where $C_\mathrm{a}$ is the added mass coefficient (in 2D $C_\mathrm{a}=1$) and $\gamma_2$ is another phenomenological adjustable parameter, due to the non-sharpness of the impurity boundary. The effective equation of motion for the impurity is then:
\begin{equation}
M_\mathrm{p} \ddot{\q} = \gamma_2 C_\mathrm{a}M_0\left( \frac{\mathrm{d}\v_\mathrm{w}}{\mathrm{d}t}\bigg\rvert_\mathbf{q} - \ddot{\q} \right) + \gamma_1M_0\frac{\mathrm{d}\v_\mathrm{w}}{\mathrm{d}t}\bigg\rvert_\mathbf{q} 
\label{Eq:Eqm}
\end{equation}
which implies
\begin{equation}
\ddot{q} = \epsilon \frac{\omega^2}{k}\sin{(kq-\omega t)},\qquad\quad \mathrm{with} \quad
\epsilon =  \eta\frac{A_\rho}{\rho_0}\quad\qquad \mathrm{and} \quad
\eta = \left(\frac{\gamma_2C_\mathrm{a}+ \gamma_1}{\gamma_2 C_\mathrm{a} + \mathcal{M}}\right)
\label{Eq:PartEqWave}
\end{equation}
Note that the contribution of the quantum pressure in (\ref{Eq:Part3}) is negligible. 
Indeed  $\nabla{\left( \frac{p_\mathrm{q}}{\rho} \right) = \frac{1}{2} \left(\frac{c^2k^2}{\omega^2}  \xi^2k^2\right) \frac{A_\rho}{\rho_0} \frac{\omega^2}{k} }$ and for long waves $\xi k \ll 1$ it can be safely neglected.

The values $\gamma_1\simeq 0.69$ and $\gamma_2\simeq 0.25$ used in the Letter were obtained by fitting subsets of the simulation data. 
Specifically, we first set $\gamma_1=1$ as for an impurity with sharp boundaries. 
Then, we obtained $\gamma_2=0.25$ by fitting the leading of our prediction using the measured drift velocity for the lightest impurity ($M=0.1$), smallest wavelength ($32\xi$) and smallest amplitude ($A=0.01$). 
Finally, once $\gamma_2$ was fixed, we made $\gamma_1$ free and determined its value by minimizing the root-mean-square error between the (normalized) displacement curves for the two extreme masses ($M=0.1$ and $M=2$ in Fig.4c of the main text).

\section{Multi-scale expansion for the evaluation of the drift}
By rescaling the variables $\tilde{q}= kq$ and $\tilde{t}= \omega t$, equation (\ref{Eq:PartEqWave}) reads
\begin{equation}
\ddot{\tilde{q}}(\tilde{t}) = \epsilon\sin{(\tilde{q}(\tilde{t}) - \tilde{t})}.
\label{Eq:PartEqWaveAd}
\end{equation}
In the following we drop the tilde for simplicity in the notation.
We use a multi-scale expansion to establish the behaviour of $q(t)$ at long times rescaled by $\epsilon^{-1}$. 
We look for a solution $q(t,\epsilon)=Q(t,\epsilon t, \epsilon)$, where $Q(t,\tau, \epsilon)$ is a function of the fast time variable $ t $ and the slow time variable $\tau$ that gives $q(t,\epsilon)$ when $\tau=\epsilon t$. Equation (\ref{Eq:PartEqWaveAd}) becomes:
\begin{equation}
\frac{\partial^2Q}{\partial t^2} + 2\epsilon\frac{\partial^2Q}{\partial t \partial \tau} + \epsilon^2\frac{\partial^2Q}{\partial \tau^2} = \epsilon\sin{(Q - t)}.
\label{Eq:Expansion}
\end{equation}
Plugging the expansion $Q(t,\tau,\epsilon)=q_0(t,\tau)+\epsilon q_1(t,\tau) + \epsilon^2 q_2(t,\tau)+\mathcal{O}(\epsilon^3)$ into (\ref{Eq:Expansion}) we obtain the hierarchy:
\begin{eqnarray}
\frac{\partial^2q_0}{\partial t^2} &=& 0\qquad\qquad\mathcal{O}(\epsilon^0),
\label{Eq:e0} \\
\frac{\partial^2q_1}{\partial t^2} + 2\frac{\partial^2q_0}{\partial t\partial \tau} &=& \sin{(q_0-t)} \qquad\qquad\mathcal{O}(\epsilon^1),
\label{Eq:e1} \\
\frac{\partial^2q_2}{\partial t^2} + 2\frac{\partial^2q_1}{\partial t\partial \tau} + \frac{\partial^2q_0}{\partial \tau^2}  &=& q_1\cos{(q_0-t)} \qquad\qquad\mathcal{O}(\epsilon^2),
\label{Eq:e2} \\
\frac{\partial^2q_3}{\partial t^2} + 2\frac{\partial^2q_2}{\partial t\partial \tau} + \frac{\partial^2q_1}{\partial \tau^2}  &=& - q_2\sin{(q_0-t)} \qquad\qquad\mathcal{O}(\epsilon^3).
\label{Eq:e3}
\end{eqnarray} 
The order $\mathcal{O}(\epsilon^0)$ implies $q_0(t,\tau)=C_{00}(\tau)t + C_{01}(\tau)$, with $C_{00}=0$ in order to cancel the secular term in the fast variable $t$. The order $\mathcal{O}(\epsilon^1)$ gives $q_1(t,\tau)=-\sin{(q_0(\tau)-t)} + C_{10}(\tau)t + C_{11}(\tau)$, and also $C_{10}(\tau)$ must vanish. Substituting these results in the equation (\ref{Eq:e2}), we obtain:
\begin{equation}
\frac{\partial^2q_2}{\partial t^2} -2\sin{(q_0-t)}\frac{\mathrm{d}q_0}{\mathrm{d}\tau} + \frac{\mathrm{d}^2q_0}{\mathrm{d}\tau^2}  = -\frac{1}{2}\sin{(2(q_0-t))}+C_{11}\cos{(q_0-t)}.
\label{Eq:e2exp}
\end{equation} 
The solution $q_2(t,\tau)$ will not contain secular terms in $t$ only if $\mathrm{d}^2q_0/\mathrm{d}\tau^2=0$, which implies
\begin{equation}
q_0(\tau)=B_1\tau + B_2,
\label{Eq:q0}
\end{equation} 
where $B_1$ and $B_2$ are constant at all time scales. Therefore, the solution at order $\mathcal{O}(\epsilon^2)$ is
\begin{equation}
q_2(t,\tau) = -2B_1\sin{(q_0(\tau)-t)} + \frac{1}{8}\sin{(2(q_0(\tau)-t))} - C_{11}(\tau)\cos{(q_0(\tau)-t)}+C_{21}(\tau).
\label{Eq:q2}
\end{equation} 
Substituting this in the equation (\ref{Eq:e3}), we obtain:
\begin{eqnarray}
\frac{\partial^2q_3}{\partial t^2} +\left( -2\frac{\mathrm{d}C_{11}}{\mathrm{d}t} -3B_1^2 + C_{21} \right)\sin{(q_0-t)} + \left( B_1+\frac{1}{8}\sin{(q_0-t)}-\frac{C_{11}}{2} \right)\sin{(2(q_0-t))} = 2B_1\sin^2{(q_0-t)} - \frac{\mathrm{d}^2C_{11}}{\mathrm{d}\tau^2}.\nonumber\\
\label{Eq:e3exp}
\end{eqnarray} 
The solution $q_3(t,\tau)$ will not contain secular terms in $t$ only if it is a periodic function of $t$. This implies that the average over $t$ of each derivative $\frac{\partial^nq_3}{\partial t^n}$ must vanish. Therefore, averaging Eq. (\ref{Eq:e3exp}) over $t$ and imposing $\left\langle \frac{\partial^2 q_3}{\partial t^2} \right\rangle_t=0$ we have the condition $\frac{\mathrm{d}^2C_{11}}{\mathrm{d}\tau^2}=B_1$, which implies
\begin{equation}
C_{11}(\tau) = \frac{B_1}{2}\tau^2 + D_1\tau + D_2,
\label{Eq:C11}
\end{equation}
where $D_1$ and $D_2$ are constant at all time scales. Therefore, the solution at order $\mathcal{O}(\epsilon^1)$ is
\begin{equation}
q=B_1\tau + B_2 + \epsilon\left[-\sin{(q_0-t)} + \frac{B_1}{2}\tau^2 + D_1\tau + D_2\right] + \mathcal{O}(\epsilon^2).
\label{Eq:qsole1}
\end{equation} 
The velocity at order $\mathcal{O}(\epsilon^2)$ is:
\begin{eqnarray}
&&\dot{q} = \frac{\partial q_0}{\partial t} + \epsilon\left(\frac{\partial q_1}{\partial t} + \frac{\partial q_0}{\partial \tau} \right)+ \epsilon^2\left(\frac{\partial q_2}{\partial t} + \frac{\partial q_1}{\partial \tau} \right) =  \nonumber \\
&&\epsilon\left[ B_1 + \cos{(q_0-t)} \right] + \nonumber \\
&&\epsilon^2\left[ B_1\cos{(q_0-t)} -\frac{1}{4}\cos{(2(q_0-t))} - \left( \frac{B_1}{2}\tau^2 + D_1\tau + D_2\right)\sin{(q_0-t)} + B_1\tau + D_1 \right]  + \mathcal{O}(\epsilon^3). \nonumber \\
\label{Eq:qdotsole1}
\end{eqnarray} 
The constants can be fixed from the initial condition. From Eq. (\ref{Eq:qsole1}) we have:
\begin{eqnarray}
B_2 = q(0),\\
D_2 = \sin{(q(0))}.
\label{Eq:posconst}
\end{eqnarray}
From Eq. (\ref{Eq:qdotsole1}) and setting $\dot{q}(0)=0$ for simplicity we have:
\begin{eqnarray}
B_1 = - \cos{(q(0))},\\
D_1 = 1 + \frac{1}{4}\cos{(2q(0))}.
\label{Eq:velconst}
\end{eqnarray}
The drift velocity is finally obtained averaging (\ref{Eq:qdotsole1}) over the fast timescale
\begin{equation}
v_\mathrm{drift} = \left\langle \dot{q}\right\rangle_t =   - \epsilon\cos(q(0)) + \epsilon^2\left( 1 + \frac{1}{4}\cos{(2q(0))} \right) + \mathcal{O}(\epsilon^3).
\label{Eq:drift}
\end{equation}

Now we can come back to the original variables and define $\varphi = kq(0)$ the the initial impurity-wave phase, with the convention that $ \varphi=0 $ when $ q(0) $ is at the wave crest. 
Therefore, the drift becomes
\begin{equation}
v_\mathrm{drift} \approx  - \frac{\omega}{k}\epsilon\cos(\varphi) + \frac{\omega}{k}\epsilon^2\left( 1 + \frac{1}{4}\cos{(2\varphi)} \right) .
\label{Eq:drift2}
\end{equation}
Notice that, even if the initial velocity of the impurity is zero, at the leading order there is still a drift proportional to the parameter $\epsilon$ that depends on the initial position of the impurity with respect to the wave. 
In particular, if the impurity is initially placed in a trough of the density wave, it will have a drift in the same verse of the wave, while if it is placed in the peak it will move backwards. 

If we average over random initial conditions, we see that there is still a net drift at the second order, that is independent on the initial position of the impurity:
\begin{equation}
\left\langle v_\mathrm{drift} \right\rangle_\varphi \approx  \frac{\omega}{k}\epsilon^2.
\label{Eq:drift_ave}
\end{equation}
Note that the same effect is recovered placing the impurity at an initial phase $\varphi = \left(2n+1\right)\pi/2$. The only difference is that the initial condition dependence of the second order does not vanish, resulting in a different prefactor:
\begin{equation}
v_\mathrm{drift}(\varphi=\pi/2) \approx  \frac{\omega}{k}\epsilon^2\left( 1 - \frac{1}{4} \right).
\label{Eq:drift_2nd_ic}
\end{equation}

It is interesting that the phase averaged drift (\ref{Eq:drift_ave}) resembles, up to a constant, to the drift $v_\mathrm{drift}^\mathrm{tracer}$ obtained for a perfect tracer (see next section).

\section{Stokes drift for a perfect tracer in a classical viscous fluid}

In a classical fluid, perfect (small) impurity tracers follow perfectly the flow under the assumption of infinite Stokes drag.
The dynamic of such impurity tracer would then be
\begin{equation}
\dot{q}=v_\mathrm{w}(q(t),t) = \varepsilon\frac{\omega}{k}\cos{(kq-\omega t)},\qquad\quad \varepsilon=\frac{A_\rho}{\rho_0}
\label{Eq:vlag}
\end{equation}
For completness, we derive here the Stokes drift in the case of a tracer in a classical fluid. 

Using the rescaled variables $\tilde{q}=kq$ and $\tilde{t}=\omega t$, equation (\ref{Eq:vlag}) reads:
\begin{equation}
\dot{\tilde{q}}(\tilde{t})=\varepsilon\cos{(\tilde{q}(\tilde{t})- \tilde{t})}.
\label{Eq:vlag_adim}
\end{equation}
This can be treated again using a multi-scale approach in the small parameter $\varepsilon$. Dropping the tilde for simplicity, we look for a solution $q(t,\varepsilon)=Q(t,\varepsilon t, \varepsilon)$, where $Q(t,\tau, \varepsilon)$ is a function of the fast time variable $t$ and the slow time variable $\tau$ that gives $q(t,\varepsilon)$ when $\tau=\varepsilon t$. Equation (\ref{Eq:vlag_adim}) becomes:
\begin{equation}
\frac{\partial Q}{\partial t} + \varepsilon\frac{\partial Q}{\partial\tau}= \varepsilon\cos{(Q - t)}.
\label{Eq:vlag_exp}
\end{equation}
Plugging the expansion $\tilde{q}(t,\tau,\varepsilon)=q_0(t,\tau)+\varepsilon q_1(t.\tau) + \varepsilon^2 q_2(t,\tau)+\mathcal{O}(\varepsilon^3)$ inside (\ref{Eq:Expansion}) we obtain the hierarchy:
\begin{eqnarray}
\frac{\partial q_0}{\partial t} &=& 0\qquad\qquad\mathcal{O}(\epsilon^0), \\
\label{Eq:el0}
\frac{\partial q_1}{\partial t} + \frac{\partial q_0}{\partial \tau} &=& \cos{(q_0-t)} \qquad\qquad\mathcal{O}(\epsilon^1), \\
\label{Eq:el1}
\frac{\partial q_2}{\partial t} + \frac{\partial q_1}{\partial \tau}  &=& -q_1\sin{(q_0-t)} \qquad\qquad\mathcal{O}(\epsilon^2). 
\label{Eq:el2}
\end{eqnarray} 
The order $\mathcal{O}(\epsilon^0)$ implies $q_0(t,\tau)=q_0(\tau)$. The order $\mathcal{O}(\epsilon^1)$ gives $q_1(t,\tau)=-\sin{(q_0(\tau)-t)} + \frac{\mathrm{d} q_0}{\mathrm{d}\tau}t + C_1(\tau)$, therefore $\frac{\mathrm{d} q_0}{\mathrm{d}\tau}=0$ in order to cancel the secular term in the fast variable. We have $q_0$ constant at all timescales, and the order $\mathcal{O}(\epsilon^2)$ becomes:
\begin{equation}
\frac{\partial q_2}{\partial t} + \frac{\mathrm{d} C_1}{\mathrm{d}\tau} = \sin^2{(q_0-t)}  - \sin(q_0-t).
\label{Eq:qlag_ord2}
\end{equation}
The function $q_2(t,\tau)$ is periodic in $t$, and therefore without secular terms in $t$, only if $\left\langle \frac{\partial q_2}{\partial t} \right\rangle_t = 0 $. Averaging Eq. (\ref{Eq:qlag_ord2}) over the fast variable we get:
\begin{equation}
\frac{\mathrm{d} C_1}{\mathrm{d}\tau}=\frac{1}{2}.
\label{Eq:condC1}
\end{equation}
Now we can write the solution for the velocity up to the order $\mathcal{O}(\varepsilon^2)$
\begin{equation}
\dot{q}= \frac{\partial q_0}{\partial t} + \varepsilon\left(\frac{\partial q_1}{\partial t} + \frac{\partial q_0}{\partial \tau} \right)+ \varepsilon^2\left(\frac{\partial q_2}{\partial t} + \frac{\partial q_1}{\partial \tau} \right) = \varepsilon^2\left( \frac{1}{2} -\sin{(q_0-t)} \right)
\label{Eq:sol_vel2}
\end{equation}
and obtain the Lagrangian drift velocity for the tracer
\begin{equation}
\left\langle \dot{q} \right\rangle_t = \frac{1}{2}\varepsilon^2,
\label{Eq:drift_lan}
\end{equation}
that in the dimensional variables is
\begin{equation}
v_\mathrm{drift}^\mathrm{tracer} = \frac{\omega}{2k}\left( \frac{A_\rho}{\rho_0} \right)^2.
\label{Eq:drift_lan}
\end{equation}

\bibliographystyle{apsrev4-1}
\bibliography{../PRL/references}